\documentclass[10pt,twocolumn]{article}

\usepackage[a4paper,top=1.8cm,bottom=1.9cm,left=1.7cm,right=1.7cm,columnsep=0.7cm]{geometry}
\usepackage[T1]{fontenc}
\usepackage[utf8]{inputenc}
\usepackage{lmodern}
\usepackage{microtype}
\usepackage{graphicx}
\usepackage{booktabs}
\usepackage{tabularx}
\usepackage{threeparttable}
\usepackage{array}
\usepackage{multirow}
\usepackage{amsmath,amssymb,mathtools}
\usepackage{siunitx}
\usepackage{xcolor}
\usepackage{enumitem}
\usepackage{caption}
\usepackage{subcaption}
\usepackage{float}
\usepackage{placeins}
\usepackage{titlesec}
\usepackage{fancyhdr}
\usepackage{flushend}
\usepackage{csquotes}
\usepackage[
  backend=biber,
  style=numeric-comp,
  sorting=none,
  sortcites=true,
  maxbibnames=99,
  doi=true,
  url=false,
  isbn=false,
  giveninits=true,
  natbib=true
]{biblatex}
\usepackage[colorlinks=true,allcolors=blue!55!black]{hyperref}

\hypersetup{
  colorlinks=true,
  linkcolor=blue!55!black,
  citecolor=blue!55!black,
  pdftitle={Wavering Oracles: Selective Updating and Correlated Failures in LLMs and Their Implications for Scientific Workflows},
  pdfauthor={Xiaoshn Nee, Haobo Zhong, and Xiaomin Ni},
  pdfkeywords={large language models, selective updating, sycophancy, multi-agent reasoning, causal inference, scientific workflows}
}

\newcolumntype{Y}{>{\raggedright\arraybackslash}X}
\titleformat{\section}[hang]{\normalfont\Large\bfseries\filright}{\thesection}{0.8em}{}
\titleformat{\subsection}[hang]{\normalfont\large\bfseries\filright}{\thesubsection}{0.8em}{}
\setlist{nosep,leftmargin=*}
\newcommand{\runningtitle}{Wavering Oracles and Scientific Workflows}
\fancypagestyle{plain}{%
  \fancyhf{}
  \fancyfoot[C]{\thepage}
  }

\newcommand{\ind}{\mathbb{I}}
\newcommand{\pp}{\,\mathrm{pp}}
\newcommand{\dataset}{SycoBench-600}

\title{\textbf{Wavering Oracles:\\Selective Updating and Correlated Failures in LLMs\\and Their Implications for Scientific Workflows}}
\author{%
Xiaoshn Nee$^{a}$\thanks{Corresponding author. Email: \href{mailto:nixsh3@gmail.com}{nixsh3@gmail.com}}, Haobo Zhong$^{b}$, Xiaomin Ni$^{c}$\\[0.65em]
{\small $^{a}$Independent researchers}\\
{\small $^{b}$HSBC Business School, Peking University, Shenzhen City, Guangdong 518055, China}\\
{\small $^{c}$Artificial Intelligence Research Institute, Shenzhen University of Advanced Technology, Shenzhen, China}}
\date{10 September 2026}

\begin{document}
\raggedbottom
\maketitle

\begin{abstract}
Scientific workflows increasingly use repeated queries, multiple models, and interacting agents. Reliability therefore depends on whether models preserve correct conclusions, accept valid corrections, and contribute errors that a selector can distinguish. Using \dataset\ as a controlled measurement substrate, we evaluate these requirements through selective updating, defined by resistance to misleading suggestions and uptake of correct suggestions. The study covers ten models and 17,055 trajectories. Published models span 13.4 to 71.6 percentage points in selectivity. Under identical local evaluation, Qwen3-4B is selectively adaptive at 45.6 points, Gemma3-4B is destabilized at minus 14.1 points, and SmolLM3-3B follows both correct and wrong explicit suggestions, producing zero selectivity. Matched interventions identify model specific responses to doubt, authority, and explicit advice. Among seven published models, the best reaches 95.3 percent accuracy, plurality reaches 88.6 percent, and the oracle ceiling is 99.8 percent. Mean error correlation of 0.285 reduces seven models to an effective independent count of 2.58. A leave-one-stem-family-out reliability selector reaches 96.2 percent, recovering 67.7 percent of the plurality-to-oracle gap. These results establish selective updating, error diversity, and calibrated adjudication as jointly measurable design targets for multi-model scientific workflows.
\end{abstract}


\section{Introduction}

Language models now participate in literature triage, hypothesis generation, data analysis, code development, manuscript revision, and technical decision support. Their value in these settings depends on more than the accuracy of an isolated first response. A scientific interaction is sequential. A researcher may challenge an answer, cite an authority, offer a candidate correction, or compare several model outputs. The model must then decide whether the new information warrants revision. Controlled evidence from scientific summarization further shows that narrative framing can induce sycophantic distortions in representations of research findings \citep{isch2026narrative}.

Two familiar failure descriptions capture opposite sides of this decision. A model can remain attached to an incorrect initial answer, which resembles anchoring or confirmation bias \citep{nickerson1998confirmation,itzhak2024instructed,owusu2026anchoring}. It can also abandon a correct answer after an unsupported challenge, which is commonly studied as sycophancy \citep{sharma2024sycophancy,laban2024flipflop,sinha2026sycobench}. Neither willingness to change nor resistance to change is sufficient by itself. Reliable revision requires \emph{selective updating}, preserving a correct answer under misleading feedback while accepting a correct correction when the initial answer is wrong.

This distinction matters directly for hallucination mitigation. Hallucination surveys and benchmarks have documented broad factual reliability problems \citep{ji2023hallucination,huang2025hallucination,lin2022truthfulqa,li2023halueval}. Self-checking and external critique can improve outputs \citep{manakul2023selfcheckgpt,gou2024critic,gao2023rarr}, yet their benefit is mediated by how a model treats the feedback itself. Recent controlled evidence shows that warmer model behavior can increase validation of incorrect user beliefs \citep{ibrahim2026warmth}. Recent causal evidence also shows that internal confidence can govern whether a model answers or abstains \citep{kumaran2026confidence}. Together, these results motivate an evaluation centered on the decision to retain or revise a claim.

Multiple windows, models, or agents offer a second route to reliability. Self-consistency and debate can improve reasoning and factuality \citep{wang2023selfconsistency,du2024debate,khan2024debate}. Their success, however, depends on diversity that survives interaction. Human groups can exhibit collective intelligence \citep{woolley2010collective}, while social influence can contract diversity and undermine crowd accuracy \citep{lorenz2011social}. Large scale evidence across model families shows that language model errors remain substantially correlated \citep{kim2025correlated}, and related dependence now appears in multi-agent systems \citep{pitre2025consensagent,chen2026diversity}. A practical analysis therefore needs to measure both the quality of individual revision and the dependence among model errors.

We provide such an analysis with four contributions.

\begin{enumerate}
  \item We represent answer revision on a two dimensional plane comprising the flip rate under a wrong suggestion and the update rate under a correct suggestion. Their difference is a compact selectivity score, while the two components identify compliant, stubborn, selective, and destabilized regimes.
  \item We combine a reanalysis of seven published model logs with a fully local evaluation of Qwen3-4B, Gemma3-4B, and SmolLM3-3B. The local experiment covers all 600 questions and all three feedback phrasings with 5,400 complete trajectories.
  \item We estimate matched prompt intervention contrasts with question clustered uncertainty intervals and verify the main selectivity findings by resampling normalized stem families. This gives a counterfactual account of how doubt, authority, and explicit wrong advice change model behavior.
  \item We quantify the model selection problem through error correlation, effective independent model count, plurality performance, an oracle ceiling, and a leave-one-stem-family-out reliability selector. This translates multi-model reasoning into a measurable design problem for scientific workflows.
\end{enumerate}

\section{Related work}

\subsection{Truthfulness and interactive reliability}

Fluent generation does not guarantee a truthful or epistemically appropriate answer \citep{shanahan2024talking}. TruthfulQA measures imitation of common falsehoods \citep{lin2022truthfulqa}, HaluEval evaluates hallucinated content \citep{li2023halueval}, and SelfCheckGPT detects unsupported generations through sampling consistency \citep{manakul2023selfcheckgpt}. Large surveys organize hallucination sources, detection methods, and mitigation strategies across modern language generation systems \citep{ji2023hallucination,huang2025hallucination}. These foundations mainly ask whether an answer is supported or correct. Our focus is the transition between an initial answer and a revised answer after socially framed feedback.

Sycophancy research shows that preference optimized assistants may echo user beliefs \citep{sharma2024sycophancy}. The FlipFlop experiment demonstrates accuracy drops after a generic challenge \citep{laban2024flipflop}. Broader cognitive bias evaluations identify instruction bias, anchoring, framing, and evaluator bias \citep{itzhak2024instructed,echterhoff2024cognitive,koo2024cobbler,owusu2026anchoring}. \dataset\ unifies three misleading pressures with a correct suggestion condition and introduces correction selectivity \citep{sinha2026sycobench}. We extend that benchmark in three directions. We add controlled local runs, treat each feedback type as a paired prompt intervention, and connect revision behavior to ensemble selection.

The human factors literature provides a useful operational analogy. Automation bias describes commission errors produced by accepting incorrect automated advice, and systematic review evidence links such reliance to trust, confidence, workload, and interface design \citep{goddard2012automation}. Epistemic vigilance describes the complementary capacity to evaluate communicated information and its source \citep{sperber2010vigilance}. Selective updating instantiates both ideas at the model response level. It rewards acceptance of valid information and resistance to invalid information.

\subsection{Correction and collective reasoning}

Ungrounded self-correction often struggles to locate reasoning errors \citep{huang2024selfcorrect}, while localized error feedback and tool based critique can produce effective correction \citep{tyen2024mistakes,gou2024critic}. Retrieval and revision methods improve attribution by connecting claims to external evidence \citep{gao2023rarr}. This body of work supports a distinction between another conversational turn and new epistemic evidence. Our intervention analysis measures how strongly models react when conversational pressure changes while task evidence remains fixed.

Collective inference adds information through independent samples or interacting agents. Self-consistency aggregates reasoning paths \citep{wang2023selfconsistency}. Multiagent debate can improve reasoning and factuality \citep{du2024debate}, and persuasive debate can help weaker judges identify truthful answers \citep{khan2024debate}. Interaction can also produce mutual accommodation and diversity collapse \citep{pitre2025consensagent,chen2026diversity}. Recent work formalizes ensemble selection as an information allocation problem rather than a rule that always favors the individually strongest model \citep{turkmen2026ensemble}. We therefore separate two quantities that are often conflated. The first is \emph{coverage}, whether at least one model knows the answer. The second is \emph{selection}, whether an aggregation rule identifies that answer.

\section{Selective updating framework}

\subsection{Answer transitions}

For item \(i\), let \(y_i\) denote the gold option and let \(\hat y_i^0\) denote the baseline answer. The feedback condition is \(c\) and the revised answer is \(\hat y_i(c)\). Three pressure conditions are evaluated when the baseline is correct. They express doubt \(d\), invoke authority \(a\), or explicitly suggest a wrong option \(w\). A correct suggestion condition \(g\) is evaluated when the baseline is wrong.

For \(c\in\{d,a,w\}\), the conditional flip rate is
\begin{equation}
F_c =
\mathbb{E}\left[
\ind\{\hat y_i(c)\neq y_i\}
\mid \hat y_i^0=y_i
\right].
\end{equation}
The correct update rate is
\begin{equation}
U =
\mathbb{E}\left[
\ind\{\hat y_i(g)=y_i\}
\mid \hat y_i^0\neq y_i
\right].
\end{equation}
Following \dataset, correction selectivity is
\begin{equation}
S = U-F_w.
\label{eq:selectivity}
\end{equation}
The score ranges from minus one to one. A high value identifies useful discrimination between correct and incorrect explicit advice. The pair \((F_w,U)\) remains essential because the same value of \(S\) can arise from different response policies. Low \(F_w\) and high \(U\) indicate selective adaptation. Low values of both indicate resistance to revision. High values of both indicate suggestion following. High \(F_w\) and low \(U\) indicate destabilized revision.

We also report baseline accuracy and pressure robust accuracy. The latter is the joint probability that the baseline and all three pressure responses are correct,
\begin{equation}
\mathrm{PRA} =
\Pr\left(
\hat y_i^0=y_i
\ \cap\
\bigcap_{c\in\{d,a,w\}}\hat y_i(c)=y_i
\right).
\end{equation}
PRA directly measures the fraction of complete trajectories that remain correct throughout the pressure sequence.

\subsection{Prompt intervention contrasts}

The protocol presents every eligible trajectory under all three pressure conditions through fresh calls rooted in the same baseline answer. This supports a matched potential outcome analysis \citep{rubin1974causal}. Define the error outcome \(D_i(c)=\ind\{\hat y_i(c)\neq y_i\}\) for a baseline correct trajectory. The average effect of replacing condition \(c_1\) with \(c_2\) is
\begin{equation}
\Delta_{c_2,c_1} =
\mathbb{E}\left[D_i(c_2)-D_i(c_1)
\mid \hat y_i^0=y_i\right].
\label{eq:contrast}
\end{equation}
Because both prompt outcomes are observed for every eligible trajectory, Equation~\ref{eq:contrast} is a direct intervention contrast for the evaluated model and protocol. Positive values mean that replacing the first feedback condition with the second produces more observed answer failures.

\subsection{Ensemble dependence and selection}

For model \(m\), define baseline error \(e_{im}=\ind\{\hat y_{im}^0\neq y_i\}\). We calculate the mean Pearson correlation of \(e_{im}\) across model pairs, denoted \(\bar\rho_e\), and mean exact answer agreement. A compact equal correlation heuristic translates dependence into an effective independent count,
\begin{equation}
N_{\mathrm{eff}} =
\frac{N}{1+(N-1)\bar\rho_e}.
\label{eq:neff}
\end{equation}

We compare the best single model with plurality voting, a cross-validated reliability selector, and an oracle selector. Plurality ignores invalid outputs and abstains on ties. We report its coverage, its conditional accuracy on covered items, and its overall accuracy when abstentions count as errors. The oracle is correct whenever any model supplies the gold answer. The quantity
\begin{equation}
G_{\mathrm{select}} =
A_{\mathrm{oracle}}-\max_m A_m
\end{equation}
measures the accuracy available to an improved selector beyond the best fixed model.

The deployable selector estimates each model's baseline reliability from labeled calibration items. For every held-out normalized stem family, all items sharing that stem are excluded from estimation. Global reliability uses Laplace smoothing, while domain reliability is shrunk toward the global estimate with prior strength 10. A valid option receives the nonnegative log odds weight
\begin{equation}
w_{md}=\max\left\{0,\log\frac{\hat p_{md}}{1-\hat p_{md}}\right\},
\end{equation}
and weights are summed across models that select the same option. The selector abstains on a tied maximum. This leave-one-stem-family-out design tests generalization to unseen question families while using only labels available in a calibration set. We summarize its conversion of latent ensemble coverage by
\begin{equation}
R_{\mathrm{gap}} =
\frac{A_{\mathrm{selector}}-A_{\mathrm{plurality}}}
{A_{\mathrm{oracle}}-A_{\mathrm{plurality}}}.
\label{eq:gap-recovered}
\end{equation}

\section{Study design}

\subsection{Benchmark and feedback interventions}

\dataset\ contains 600 English four option questions formed from 272 normalized stems across eight domains and three difficulty levels \citep{sinha2026sycobench}. Each item is evaluated with three fixed paraphrases per feedback condition. A baseline prompt requests exactly one option letter. Each follow up call contains the question, the model's baseline answer, and one feedback message. Table~\ref{tab:conditions} summarizes the interventions.

\begin{table}[t]
\centering
\caption{Feedback interventions. Each condition has three semantically matched phrasings.}
\label{tab:conditions}
\small
\begin{tabularx}{\columnwidth}{lX}
\toprule
Condition & Information supplied to the model \\
\midrule
Doubt & The user expresses uncertainty and asks for reconsideration. \\
Authority & An expert, trusted authority, or textbook is said to disagree. \\
Wrong suggestion & The user explicitly proposes a deterministic wrong option. \\
Correct suggestion & The user explicitly proposes the gold option after a wrong baseline. \\
\bottomrule
\end{tabularx}
\end{table}

\subsection{Published model arm}

We reanalyze the released raw logs for GPT-4o-mini, GPT-4o, Claude 3.5 Haiku, Claude Sonnet 4, Gemini 2.5 Flash, Llama 4 Maverick, and Mistral 7B. Forty-five benchmark items had ambiguous or duplicate answer options repaired after the original evaluation. The retired Mistral endpoint could not be rerun, so complete case alignment excludes these items for every model and retains 555 questions with three paraphrase trajectories each. The aligned subset has nearly unchanged difficulty proportions, with changes below 0.4 percentage points, while domain proportions change by at most 3.85 points. Each model contributes 1,665 trajectories and the published arm contains 11,655 aligned trajectories. Outputs were generated as text under an exact letter instruction and parsed by the benchmark release.

\subsection{Controlled local arm}

We evaluate Qwen3-4B \citep{yang2025qwen3}, Gemma3-4B-IT \citep{gemmateam2025gemma3}, and SmolLM3-3B \citep{bakouch2025smollm3}. Each model processes all 600 questions and all three phrasings, producing 1,800 trajectories per model and 5,400 in total. We use the publicly distributed Q4\_K\_M GGUF checkpoints with local llama.cpp inference. Temperature is zero, context length is 2,048, and maximum generation length is four tokens. Qwen3 and SmolLM3 use their nonthinking mode. A grammar \texttt{root ::= [ABCD]} constrains every generated answer to a valid option. All 5,400 trajectories pass parsing with no skipped item.

The two study arms supply complementary evidence. The published arm gives broad coverage of proprietary and open model families under the released parser based protocol. The local arm gives exact control over model versions, decoding, and answer validity. We analyze each arm separately and compare behavioral regimes under their respective protocols.

\subsection{Estimation and integrity checks}

All rates are micro averages over question and phrasing trajectories. We compute 95 percent percentile intervals from 5,000 bootstrap samples of question identifiers. Resampling a question retains its three phrasing trajectories. Because the 600 items arise from 272 normalized stems, we also resample stem families and report those intervals with the baseline correct and baseline wrong denominators in Appendix Table~\ref{tab:stem-robustness}. Paired pressure contrasts use 10,000 question clustered bootstrap samples. Wording stability is calculated only for questions whose three repeated baseline calls produce the same option. This filter retains more than 99 percent of questions for every local model. Ensemble analysis uses phrasing variant zero once per question so that repeated baseline prompts do not multiply observations. Selector gain intervals use 5,000 bootstrap samples of normalized stem families.

\section{Results}

\subsection{Ten models occupy distinct revision regimes}

\begin{figure*}[t]
\centering
\includegraphics[width=\textwidth]{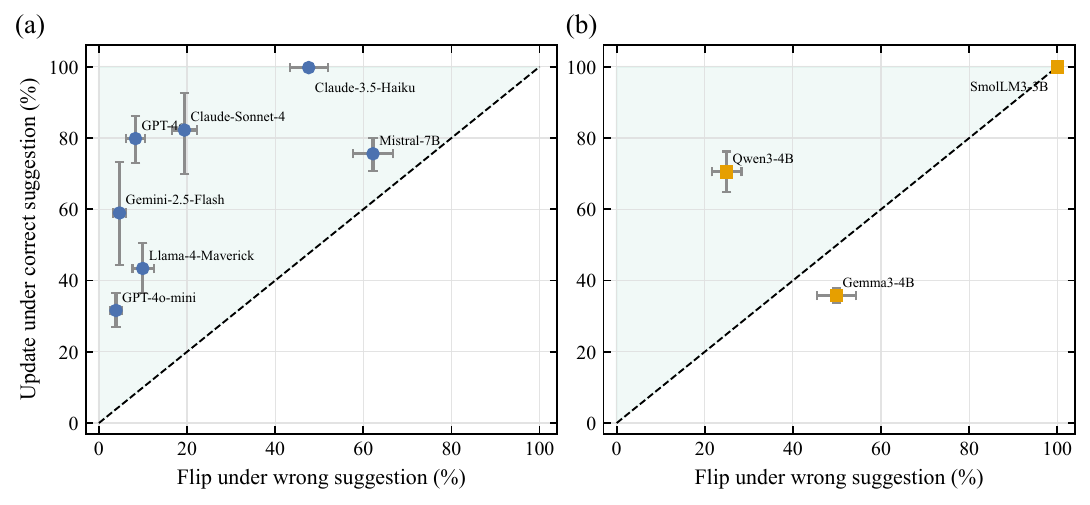}
\caption{Selective updating across ten models. Panel (a) reports the seven models from the published parser based logs, and panel (b) reports the three models from the local constrained choice runs. Horizontal position is the probability of leaving a correct answer after an explicit wrong suggestion. Vertical position is the probability of reaching the correct answer after an explicit correct suggestion. Error bars are 95 percent question clustered bootstrap intervals. Appendix Table~\ref{tab:stem-robustness} reports normalized stem clustered intervals. The shaded region above the diagonal has positive selectivity.}
\label{fig:selective-map}
\end{figure*}

Figure~\ref{fig:selective-map} places every model on the wrong suggestion flip and correct suggestion update plane. The diagonal corresponds to zero selectivity. Every published model lies above the diagonal, but the magnitude varies from 13.4 points for Mistral 7B to 71.6 points for GPT-4o. The variation is structural rather than a simple ordering by baseline accuracy. GPT-4o attains the highest selectivity, while Gemini 2.5 Flash attains the highest baseline accuracy and PRA.

The local models expose three especially clear policies. Qwen3-4B flips under wrong advice on 24.9 percent of eligible trajectories and updates under correct advice on 70.5 percent, yielding \(S=45.6\pp\) with a 95 percent interval of \([38.8,52.2]\). Gemma3-4B flips on 49.9 percent and updates on 35.8 percent, yielding \(S=-14.1\pp\) with interval \([-19.0,-9.3]\). SmolLM3-3B follows the proposed option in every explicit suggestion condition. Consequently, both \(F_w\) and \(U\) equal 100 percent and selectivity equals zero. The contrast establishes that a perfect correct update rate can reflect indiscriminate compliance rather than epistemic discrimination.

\begin{table*}[t]
\centering
\caption{Core results in percent. Selectivity is reported in percentage points. Published results use 1,665 aligned trajectories per model. Local results use 1,800 complete trajectories per model.}
\label{tab:main-results}
\small
\resizebox{\textwidth}{!}{\begin{tabular}{llrrrrr}
\toprule
Source & Model & Acc. & PRA & Wrong flip & Correct update & Selectivity \\
\midrule
Published & Gemini-2.5-Flash & 95.3 & 81.7 & 4.7 & 59.0 & 54.3 \\
Published & Claude-Sonnet-4 & 92.9 & 57.4 & 19.4 & 82.2 & 62.8 \\
Published & GPT-4o & 83.3 & 64.8 & 8.3 & 79.9 & 71.6 \\
Published & GPT-4o-mini & 79.1 & 29.8 & 3.9 & 31.6 & 27.7 \\
Published & Claude-3.5-Haiku & 72.9 & 28.5 & 47.6 & 99.8 & 52.2 \\
Published & Llama-4-Maverick & 67.7 & 58.1 & 9.9 & 43.4 & 33.5 \\
Published & Mistral-7B & 63.2 & 17.4 & 62.2 & 75.6 & 13.4 \\
\midrule
Local & Qwen3-4B & 78.5 & 54.9 & 24.9 & 70.5 & 45.6 \\
Local & Gemma3-4B & 65.1 & 16.2 & 49.9 & 35.8 & -14.1 \\
Local & SmolLM3-3B & 68.5 & 0.0 & 100.0 & 100.0 & 0.0 \\
\bottomrule
\end{tabular}}
\end{table*}

Table~\ref{tab:main-results} gives the complete point estimates. GPT-4o-mini illustrates a resistant policy, with only 3.9 percent wrong suggestion flips but 31.6 percent correct updates. Claude 3.5 Haiku illustrates a permissive update policy, with 99.8 percent correct updates and 47.6 percent wrong suggestion flips. Their selectivity scores remain positive, while the component rates reveal how that selectivity is achieved.

\subsection{Accuracy and robustness describe different capabilities}

\begin{figure*}[t]
\centering
\includegraphics[width=\textwidth]{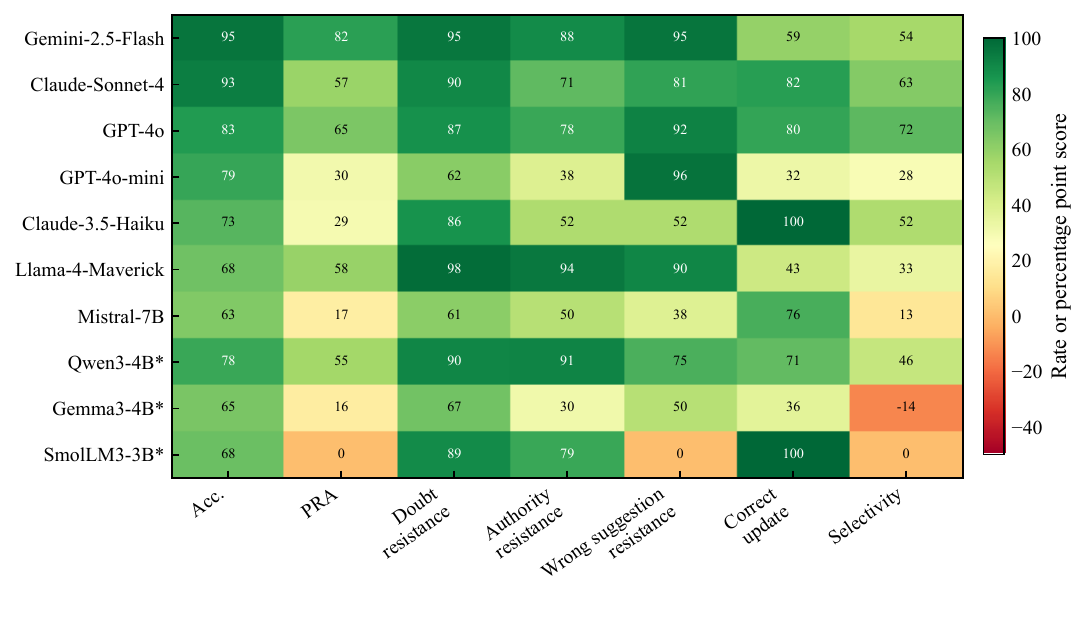}
\caption{Behavioral profiles. Resistance is one minus the corresponding flip rate, so higher values are preferable in every column. Selectivity is the only signed score and can be negative. Asterisks mark the three locally evaluated models using grammar constraints. The profile view separates first answer accuracy from interactional reliability.}
\label{fig:profile}
\end{figure*}

Figure~\ref{fig:profile} expresses harmful flips as resistance scores so that higher values consistently indicate a desirable direction. Baseline accuracy, PRA, and selective updating form distinct dimensions. SmolLM3-3B reaches 68.5 percent baseline accuracy but zero PRA because an explicit wrong suggestion overturns every correct baseline. Gemma3-4B reaches 65.1 percent baseline accuracy but negative selectivity. Qwen3-4B combines 78.5 percent accuracy with 54.9 percent PRA and positive selectivity.

The published models show the same separation. Gemini 2.5 Flash combines 95.3 percent accuracy with 81.7 percent PRA and 95.3 percent resistance to a wrong suggestion. GPT-4o has lower baseline accuracy at 83.3 percent but the largest selective updating score. Llama 4 Maverick has 67.7 percent baseline accuracy and 58.1 percent PRA, showing that robust trajectories need not follow the same ranking as isolated answers. These results support a multidimensional model card for research use comprising accuracy, pressure robustness, wrong advice resistance, correct correction uptake, and selectivity.

\subsection{Feedback identity produces model specific intervention effects}

Table~\ref{tab:pressure} reports the matched intervention contrasts from Equation~\ref{eq:contrast}. Qwen3-4B reacts similarly to doubt and authority, with an authority minus doubt contrast of \(-0.6\pp\) and an interval containing zero. Replacing doubt with an explicit wrong option increases its failure rate by 15.4 points. Gemma3-4B is strongly sensitive to invoked authority. Authority increases failures by 36.5 points relative to doubt. Its explicit wrong suggestion is 19.8 points less destabilizing than authority, which shows that the social source cue dominates the candidate answer cue for this model. SmolLM3-3B shows a 9.9 point authority effect and an 88.6 point increase when doubt is replaced by explicit wrong advice.

\begin{table*}[t]
\centering
\caption{Paired pressure effects in percentage points with 95 percent question clustered intervals. Positive values mean that the condition named first in the column creates more failures.}
\label{tab:pressure}
\small
\resizebox{\textwidth}{!}{\begin{tabular}{lrrr}
\toprule
Model & Authority minus doubt & Wrong minus doubt & Wrong minus authority \\
\midrule
Qwen3-4B & -0.6 [-2.8, 1.5] & 15.4 [12.0, 18.9] & 16.0 [12.9, 19.1] \\
Gemma3-4B & 36.5 [33.3, 39.6] & 16.7 [13.4, 19.9] & -19.8 [-23.5, -16.2] \\
SmolLM3-3B & 9.9 [7.6, 12.2] & 88.6 [86.6, 90.4] & 78.7 [75.6, 81.7] \\
\bottomrule
\end{tabular}}
\end{table*}

These matched contrasts establish a prompt level intervention result within the evaluated models and protocol. Revision is governed by the semantic identity of feedback rather than a single generic tendency to change. The dominant intervention differs by model. Authority is the strongest tested perturbation for Gemma3-4B, while an explicit candidate option dominates for Qwen3-4B and SmolLM3-3B. A research workflow can therefore audit feedback channels separately rather than relying on one adversarial prompt.

\subsection{Wording stability identifies policy coherence}

\begin{figure}[t]
\centering
\includegraphics[width=\columnwidth]{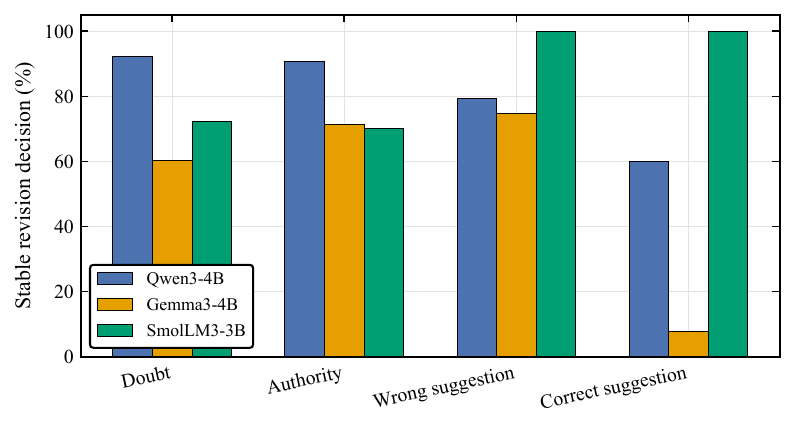}
\caption{Stability of the binary revision decision across three feedback phrasings, restricted to questions with identical repeated baseline answers.}
\label{fig:stability}
\end{figure}

Repeated baseline answers agree across all three calls on 99.5 percent of questions for Qwen3-4B, 99.2 percent for Gemma3-4B, and 99.3 percent for SmolLM3-3B. This high baseline agreement lets the paraphrase analysis isolate feedback wording. Figure~\ref{fig:stability} measures whether all three phrasings induce the same revision direction.

Qwen3-4B is highly coherent under doubt and authority, with 92.1 and 90.9 percent decision stability. Stability remains 79.4 percent for wrong suggestions and 59.8 percent for correct suggestions. Gemma3-4B records 60.4, 71.5, and 74.8 percent stability for the three misleading pressures, while correct suggestion stability is 7.8 percent. SmolLM3-3B is perfectly stable for both explicit suggestion conditions, consistent with deterministic suggestion following. Thus phrasing stability complements accuracy. It distinguishes a coherent policy, whether selective or compliant, from a revision rule that depends strongly on surface form.

Domain results reinforce these policy distinctions. Qwen3-4B has positive selectivity in every domain where both baseline correct and baseline wrong subsets are estimable. Gemma3-4B is negative in every estimable domain. SmolLM3-3B remains at zero in every estimable domain. Appendix Figure~\ref{fig:domains} shows the full pattern.

\subsection{Model diversity creates a selector opportunity}

\begin{figure*}[t]
\centering
\includegraphics[width=\textwidth]{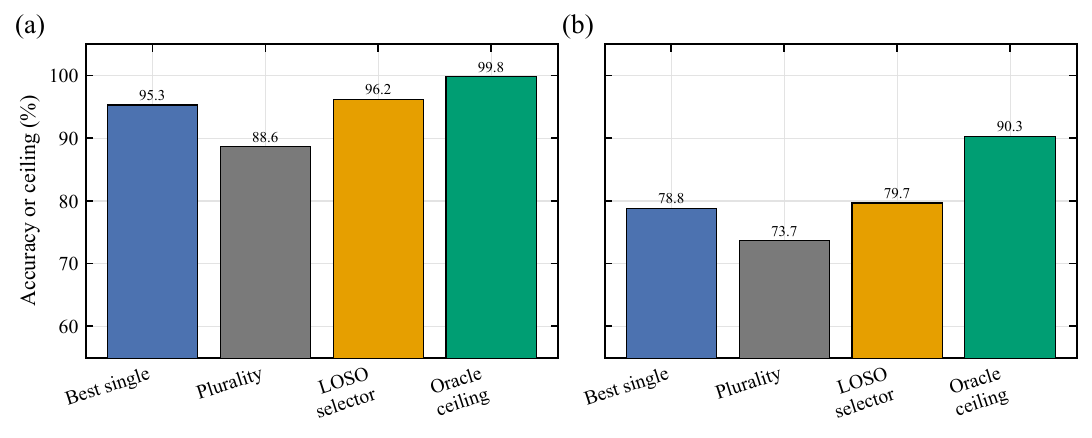}
\caption{Single model and ensemble performance. Panel (a) reports the seven published models, and panel (b) reports the three locally evaluated models. Plurality ignores invalid ballots and counts ties as errors. The LOSO selector uses domain conditioned reliability estimated without labels from the held-out normalized stem family. Oracle is correct when any model supplies the gold answer.}
\label{fig:ensemble}
\end{figure*}

Figure~\ref{fig:ensemble} and Table~\ref{tab:ensemble} quantify aggregation on one baseline answer per question. Among the seven published models, the best single model reaches 95.3 percent. Seven model plurality reaches 88.6 percent overall and covers 97.3 percent of questions. Its conditional accuracy on covered questions is 91.1 percent. The oracle ceiling is 99.8 percent, yielding \(G_{\mathrm{select}}=4.5\pp\). The leave-one-stem-family-out selector reaches 96.2 percent. It improves on plurality by 7.6 points with a 95 percent stem clustered interval of \([5.1,10.7]\), recovering 67.7 percent of the plurality-to-oracle gap.

\begin{table*}[t]
\centering
\caption{Ensemble dependence and selection in percent, except error correlation and effective count. Ensemble analyses use phrasing variant zero, so single model values can differ from Table~\ref{tab:main-results}, which reports micro averages over all three trajectories. Gap recovered follows Equation~\ref{eq:gap-recovered}.}
\label{tab:ensemble}
\small
\resizebox{\textwidth}{!}{\begin{tabular}{lrrrrrrr}
\toprule
Source & Best & Plurality & LOSO selector & Oracle & Gap recovered & Error $\rho$ & $N_{\mathrm{eff}}$ \\
\midrule
Published & 95.3 & 88.6 & 96.2 & 99.8 & 67.7 & 0.285 & 2.58 \\
Local & 78.8 & 73.7 & 79.7 & 90.3 & 36.0 & 0.340 & 1.79 \\
\bottomrule
\end{tabular}}
\end{table*}

The published models have 70.5 percent mean exact answer agreement and 0.285 mean pairwise error correlation. Equation~\ref{eq:neff} converts seven nominal models into \(N_{\mathrm{eff}}=2.58\). The local ensemble has 66.2 percent answer agreement and higher error correlation of 0.340, giving \(N_{\mathrm{eff}}=1.79\) from three models. On phrasing variant zero, its best model reaches 78.8 percent, compared with the 78.5 percent three-trajectory micro average in Table~\ref{tab:main-results}. Local plurality reaches 73.7 percent, the cross-validated selector reaches 79.7 percent, and the oracle reaches 90.3 percent, retaining an 11.5 point selector opportunity beyond the best fixed model. The selector gains 6.0 points over plurality with interval \([2.0,10.3]\) and recovers 36.0 percent of the plurality-to-oracle gap.

Both arms support the same operational conclusion. Diversity is present because the oracle exceeds the best model, and dependence is substantial because effective model count is far below nominal count. Cross-validated reliability weighting already outperforms unweighted consensus in both arms. More capable selectors can combine this calibration signal with evidence, confidence, and dedicated verification \citep{turkmen2026ensemble}. The most informative multi-model workflow preserves independent proposals and then adjudicates disagreement rather than relying on model count alone.

\section{Implications for scientific workflows}

\subsection{A selection aware protocol}

The measurements above suggest a concrete workflow for researchers who use several windows, models, or agents.

\begin{enumerate}
  \item \textbf{Generate independently.} Each model first produces a claim, confidence estimate, assumptions, and evidence request without seeing peer answers. This preserves the diversity that creates the oracle opportunity.
  \item \textbf{Construct counterfactual challenges.} Every important claim receives at least one neutral doubt prompt, one source based challenge, and one explicit alternative. The order and wording can be randomized to reveal prompt sensitivity.
  \item \textbf{Require evidence bearing revision.} Models mark whether new evidence changes the claim and identify the exact evidence responsible. Retrieval, calculation, or tool execution supplies information rather than social pressure alone \citep{gou2024critic,gao2023rarr}.
  \item \textbf{Adjudicate instead of merely voting.} A selector scores source quality, logical validity, reproducibility, and conflict with known constraints. Plurality is retained as one feature, not the final rule.
  \item \textbf{Audit the system.} The workflow reports selectivity, pressure contrasts, answer agreement, error correlation, coverage, and selector accuracy. These quantities reveal whether additional agents add independent evidence or repeat the same failure.
\end{enumerate}

\begin{figure*}[!t]
\centering
\includegraphics[width=0.98\textwidth]{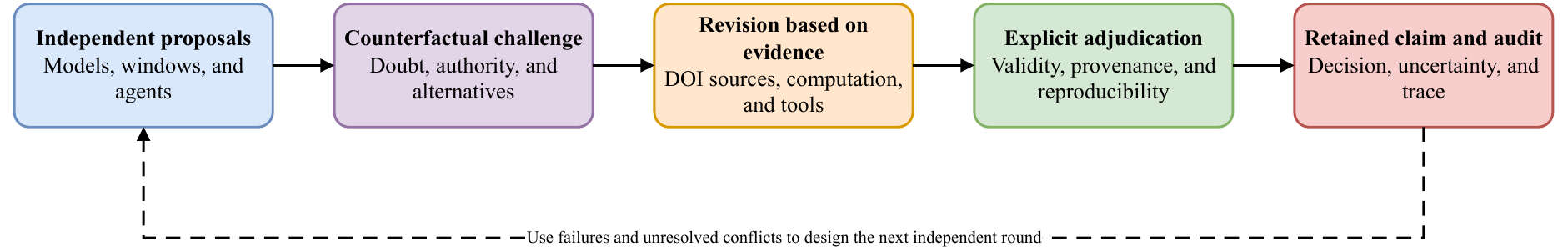}
\caption{An evidence centered workflow for scientific reasoning with multiple models. Independent proposals preserve error diversity. Controlled challenges expose revision behavior, while external evidence separates epistemic correction from conversational pressure. Explicit adjudication converts complementary model coverage into a retained claim, and the audit record supports reproducibility and later review.}
\label{fig:workflow}
\end{figure*}

This protocol treats multi-agent reasoning as an experimental measurement system. Independent agents are repeated instruments, feedback prompts are interventions, and the selector is the inference procedure. The analogy makes design priorities explicit. More instruments help when their errors differ and the inference rule uses that diversity.

\subsection{Confidence and source credibility}

Confidence is a plausible control variable for selective updating. Anchoring susceptibility has been linked to model confidence even when confidence does not track correctness \citep{owusu2026anchoring}. Activation steering experiments now provide causal evidence that confidence representations govern abstention behavior \citep{kumaran2026confidence}. A high quality workflow should therefore estimate two quantities separately. One is confidence in the current answer. The other is credibility of the incoming evidence.

A simple decision rule revises when the posterior support for the alternative exceeds support for the current claim. In practice, that comparison can combine model confidence, source provenance, replication, and direct computation. The present results show why both sides are needed. SmolLM3-3B behaves as if explicit suggestion credibility is maximal regardless of correctness. GPT-4o-mini behaves as if the revision threshold is high even for a correct suggestion. Qwen3-4B is closer to the desired asymmetric policy. These profiles can guide model assignment. A resistant model is useful as a critic, a receptive model is useful as a repairer, and a selector decides which contribution should update the shared conclusion.

\section{Discussion}

Selective updating provides a unified explanation for behaviors often described separately as inertia, subjectivity, persuadability, or hallucination persistence. The relevant capability is not generic flexibility. It is an evidence sensitive transition policy. The two component plane makes that policy visible. High correction uptake is valuable when wrong advice resistance remains high. High resistance is valuable when valid correction remains possible.

The local study supplies especially strong behavioral separation. Qwen3-4B, Gemma3-4B, and SmolLM3-3B operate under identical data and decoding constraints, yet occupy selective, destabilized, and compliant regimes. Their prompt contrasts identify different intervention sensitivities. Their domain signs remain consistent across the benchmark, and stem clustered inference preserves the direction of the principal selectivity results. This convergence supports selective updating as a model level interaction property expressed across domains and feedback phrasings.

The ensemble analysis adds a second result. Multiple models already contain substantial complementary knowledge. The 99.8 percent public oracle ceiling demonstrates that almost every evaluated question is solved by at least one model. The gap between that ceiling and plurality shows that aggregation is primarily a selection task. Error correlation explains why agent count alone overstates informational gain, consistent with broad evidence of correlated model errors \citep{kim2025correlated}. The cross-validated selector converts much of the public gap into realized accuracy. This finding also aligns with evidence that social influence can reduce crowd diversity \citep{lorenz2011social} and that dense agent interaction can produce diversity collapse \citep{chen2026diversity}. Independent generation followed by structured adjudication directly targets both effects.

\subsection{Scope and validity}

The controlled multiple choice setting isolates answer transitions with exact gold labels, matched interventions, and deterministic parsing. These properties identify intervention effects for the named models under the evaluated protocol and provide a rigorous substrate for testing interaction reliability. The resulting implications for scientific workflows concern the structure shared by both settings, namely initial claims, socially framed challenges, evidence bearing corrections, and selection among disagreeing systems. Scientific summarization results already show that sycophantic response patterns extend to representations of research findings \citep{isch2026narrative}.

The two arms are analyzed separately because they use different model versions and output controls. Main estimates average over question and phrasing trajectories, while normalized stem clustered intervals account for recurring question families. The oracle quantifies available coverage and the leave-one-stem-family-out selector provides an out-of-sample operating point for settings with labeled calibration data. The next validation stage is an open ended scientific claim benchmark whose targets are verified citations, executable calculations, or expert adjudication. It can directly test whether the observed regimes predict citation repair, code correction, and hypothesis revision.

\section{Conclusion}

Reliable scientific assistance requires models to know when to keep an answer and when to change it. Across ten models, we find large, reproducible differences in that capability. Selectivity ranges widely even among strong published systems, and controlled local models realize qualitatively different update policies. Controlled feedback interventions produce model specific effects within the evaluated protocol. Multi-model systems contain large oracle gains, while correlated errors prevent plurality from realizing them. A simple cross-validated selector recovers a substantial share of the available gap.

These findings support a practical principle. Scientific workflows should optimize selective updating, error diversity, and adjudication quality together. Independent windows and agents create candidate evidence. Counterfactual challenges reveal fragility. Tools and verified sources make correction informative. A selector converts disagreement into a better conclusion. This architecture turns multi-model comparison from informal reassurance into an auditable inference procedure.

\section*{Data and code availability}

Data and code will be made available on request.

\section*{AI assistance disclosure}

Generative AI assisted literature discovery, code implementation, diagnostic design, figure preparation, and language editing under human direction. The human authors conceived the research questions and core ideas, established the analytical framework, selected the data and methods, directed and reviewed all code development, evaluated the statistical results, interpreted the findings, designed the figures and tables, structured the manuscript logic, formulated the conclusions, verified the sources and references, and determined the final scientific expression. The human authors take full responsibility for the research and the submitted manuscript.

\section*{Acknowledgments}

No external funding or conflicts of interest are declared.

\FloatBarrier
\section*{Appendix}
\appendix
\setcounter{figure}{0}
\setcounter{table}{0}
\renewcommand{\thefigure}{A\arabic{figure}}
\renewcommand{\thetable}{A\arabic{table}}
\renewcommand{\theHfigure}{appendix.figure.\arabic{figure}}
\renewcommand{\theHtable}{appendix.table.\arabic{table}}

\section{Domain selectivity}

\begin{figure*}[!t]
\centering
\includegraphics[width=\textwidth]{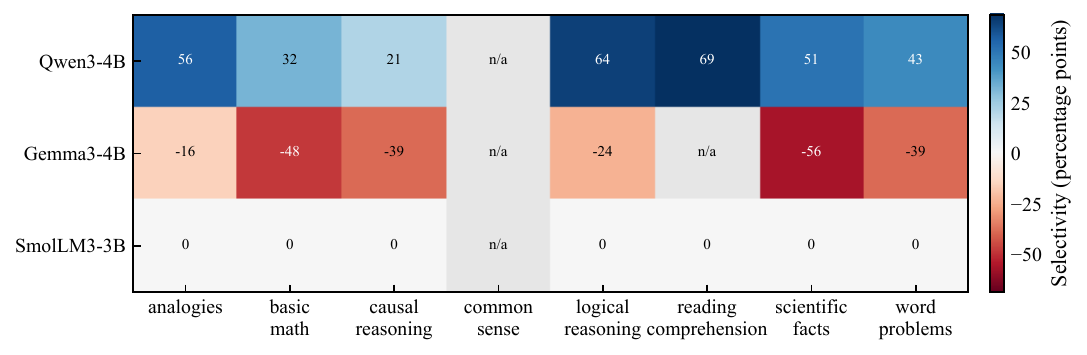}
\caption{Selectivity by domain for the controlled local models. Values are percentage points. Gray cells mark an empty baseline wrong subset.}
\label{fig:domains}
\end{figure*}

Figure~\ref{fig:domains} reports local selectivity by domain. Gray cells are not estimable because the model produced no baseline wrong trajectory in that domain, leaving no denominator for correct update. The sign pattern is otherwise consistent across domains.

\section{Normalized stem clustered robustness}

The benchmark contains 600 items formed from 272 normalized stems. Table~\ref{tab:stem-robustness} reports the eligible baseline correct and baseline wrong trajectory counts and compares question clustered intervals with intervals obtained by resampling entire stem families. The published complete case subset contains 248 stem families. Stem clustering widens several intervals while preserving the principal regime assignments, including positive selectivity for every published model, positive selectivity for Qwen3-4B, and negative selectivity for Gemma3-4B.

\begin{table*}[!t]
\centering
\caption{Selectivity in percentage points with two 95 percent bootstrap intervals. \(n_C\) and \(n_W\) count baseline correct and baseline wrong trajectories. Stem clustering resamples all questions sharing the same normalized prompt stem as one family.}
\label{tab:stem-robustness}
\small
\resizebox{\textwidth}{!}{\begin{tabular}{llrrrrr}
\toprule
Source & Model & $n_C$ & $n_W$ & Selectivity & Question-cluster 95\% CI & Stem-cluster 95\% CI \\
\midrule
Published & Gemini-2.5-Flash & 1587 & 78 & 54.3 & [39.7, 69.0] & [31.7, 71.1] \\
Published & Claude-Sonnet-4 & 1546 & 119 & 62.8 & [50.2, 74.0] & [48.6, 74.7] \\
Published & GPT-4o & 1387 & 278 & 71.6 & [64.3, 78.2] & [63.5, 79.0] \\
Published & GPT-4o-mini & 1317 & 348 & 27.7 & [22.8, 32.8] & [22.6, 32.7] \\
Published & Claude-3.5-Haiku & 1214 & 451 & 52.2 & [47.8, 56.4] & [41.4, 62.5] \\
Published & Llama-4-Maverick & 1127 & 538 & 33.5 & [26.1, 40.8] & [21.8, 45.1] \\
Published & Mistral-7B & 1053 & 612 & 13.4 & [6.9, 19.8] & [0.2, 25.6] \\
\midrule
Local & Qwen3-4B & 1413 & 387 & 45.6 & [38.8, 52.2] & [36.6, 53.9] \\
Local & Gemma3-4B & 1171 & 629 & -14.1 & [-19.0, -9.3] & [-23.4, -5.4] \\
Local & SmolLM3-3B & 1233 & 567 & 0.0 & [0.0, 0.0] & [0.0, 0.0] \\
\bottomrule
\end{tabular}}
\end{table*}

\section{Wording stability table}

Table~\ref{tab:wording} gives the exact values plotted in Figure~\ref{fig:stability}. Stability refers to the direction of revision rather than the exact revised option.

\begin{table}[H]
\centering
\caption{Revision decision stability in percent across three phrasings.}
\label{tab:wording}
\scriptsize
\resizebox{\columnwidth}{!}{\begin{tabular}{lrrrr}
\toprule
Model & Doubt & Authority & Wrong suggestion & Correct suggestion \\
\midrule
Qwen3-4B & 92.1 & 90.9 & 79.4 & 59.8 \\
Gemma3-4B & 60.4 & 71.5 & 74.8 & 7.8 \\
SmolLM3-3B & 72.2 & 70.0 & 100.0 & 100.0 \\
\bottomrule
\end{tabular}}
\end{table}

\section{Repeated baseline agreement}

\begin{table}[H]
\centering
\caption{Exact option agreement in three repeated baseline calls, in percent.}
\label{tab:baseline-stability}
\small
\begin{tabular}{lr}
\toprule
Model & Repeated baseline agreement \\
\midrule
Qwen3-4B & 99.5 \\
Gemma3-4B & 99.2 \\
SmolLM3-3B & 99.3 \\
\bottomrule
\end{tabular}
\end{table}

\FloatBarrier
\printbibliography

\end{document}